\documentclass[aps,prl,twocolumn,notitlepage,showpacs,superscriptaddress,am]{revtex4-2}%
\usepackage{graphicx}
\usepackage{amsmath}
\usepackage{amssymb}
\usepackage{mhchem}
\usepackage{color}
\usepackage{amsfonts}%
\usepackage[breaklinks=true,colorlinks=true,linkcolor=blue,urlcolor=blue,citecolor=blue]{hyperref}

\providecommand{\U}[1]{\protect\rule{.1in}{.1in}}
\makeatletter
\def\@caption@fignum@sep{\,}
\def\fnum@figure{\textbf{Fig.\thefigure}\,\textbar}
\makeatother

\begin{document}
\title{Tracking flat bands via phonon-mediated interband scattering}

\author{Fabian Garmroudi}
\email{Corresponding author: fgarmroudi@lanl.gov}
\affiliation{Materials Physics Applications--Quantum, Los Alamos National Laboratory, Los Alamos, New Mexico 87545, USA}
\author{Jennifer Coulter}
\affiliation{Center for Computational Quantum Physics, Flatiron Institute, New York 10010, USA}
\author{Illia Serhiienko}
\affiliation{Institute of Solid State Physics, TU Wien, 1040 Vienna, Austria}
\author{Christoph Fritzl}
\affiliation{Institute of Solid State Physics, TU Wien, 1040 Vienna, Austria}
\author{Xinlin Yan}
\affiliation{Institute of Solid State Physics, TU Wien, 1040 Vienna, Austria}
\author{Silke Paschen}
\affiliation{Institute of Solid State Physics, TU Wien, 1040 Vienna, Austria}
\author{Sean M. Thomas}
\affiliation{Materials Physics Applications--Quantum, Los Alamos National Laboratory, Los Alamos, New Mexico 87545, USA}
\author{Eric D. Bauer}
\affiliation{Materials Physics Applications--Quantum, Los Alamos National Laboratory, Los Alamos, New Mexico 87545, USA}
\author{Andrej Pustogow}
\affiliation{Institute of Solid State Physics, TU Wien, 1040 Vienna, Austria}
\author{Priscila F.\,S. Rosa}
\affiliation{Materials Physics Applications--Quantum, Los Alamos National Laboratory, Los Alamos, New Mexico 87545, USA}


\begin{abstract}
\noindent\textbf{	Flat-band (FB) systems offer a fertile platform for realizing exotic correlated and topological phases, yet central challenges remain in tuning FBs towards the Fermi level $E_\text{F}$ and in understanding their impact on electronic transport. Identifying how close a FB lies to $E_\text{F}$ normally requires advanced spectroscopic probes or extensive doping studies hindering fast screening of promising candidates. Here, we show that standard electrical resistivity measurements already contain a clear and universal fingerprint of the proximity of a FB to $E_\text{F}$: A characteristic curvature in the temperature-dependent resistivity at high temperatures, i.e., sublinear (superlinear) behavior when $E_\text{F}$ lies within (away from) the FB states. Within a phenomenological model, we can attribute this response to the rapid change in scattering phase space associated with interband transitions from dispersive to FB states. Notably, this characteristic emerges independently of microscopic details, such as the electron-phonon coupling strength, as confirmed by comparison with ab initio electron-phonon transport calculations. We validate this mechanism across several FB materials\,--\,including kagome, pyrochlore and Cu$_3$Au-type intermetallics\,--\,by combining high-temperature resistivity measurements with a modified Bloch-Grüneisen analysis. Our results establish a simple and broadly applicable method to locate FBs via electronic transport, enabling rapid screening of promising FB materials.}
\end{abstract}

\maketitle
\section*{Introduction}
\begin{figure}[tbh]
\centering
 \includegraphics[width=0.45\textwidth]{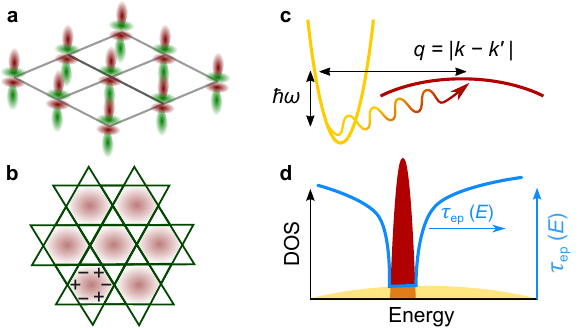}
\caption{\textbf{Sketch of flat band formation and phonon-mediated interband scattering.} Localized states and flat bands can arise in \textbf{a}, materials with localized atomic orbitals and \textbf{b}, driven by destructive phase interference patterns. \textbf{c}, Sketch of phonon-mediated interband scattering between a dispersive and a flattened band. \textbf{d}, Schematic density of states for a metallic system with a dispersive conduction band and overlapping FB. The variation of scattering phase space creates a sharp drop in the carrier relaxation time (schematically drawn as blue line).
}
\label{Fig1}
\end{figure}

\noindent Materials in which the electronic bandwidth $W$ becomes comparable to the Coulomb interaction $U$ between electrons can host a wide range of emergent phenomena, such as magnetically and charge-ordered phases, unconventional superconductivity, and fractional quantum Hall states \cite{hubbard1963electron,anderson1987resonating,steglich1979superconductivity,bednorz1986possible,tsui1982two,bolotin2009observation}. FBs are naturally found in materials with spatially localized $d$ and $f$ atomic orbitals (Fig.\,\ref{Fig1}a). However, they may also be achieved from extended wave functions through destructive phase interference of hopping paths (Fig.\,\ref{Fig1}b), as found, for instance, in the kagome lattice -- a corner sharing network of triangles \cite{zong2016observation,li2018realization,kang2020dirac}. In particular, topological FB materials, with vanishingly small $W$, have received enormous interest in recent years as they allow accessing energy scales characterized by relatively large values of $U/W$ even with moderate values of $U$ \cite{balents2020superconductivity,regnault2022catalogue,checkelsky2024flat,rosa2024quantum}. Lastly, it has been shown that FBs can be created in low-dimensional materials, such as twisted bilayer graphene, through band folding in moir\'e superlattices \cite{lisi2021observation}, leading to a variety of correlated ground states when the chemical potential is placed in the vicinity of the FBs \cite{cao2018unconventional,wong2020cascade,nuckolls2020strongly}. As a result, theoretical attempts to achieve a microscopic understanding of the physical properties of FB materials have focused on electron-electron interactions \cite{derzhko2015strongly,hu2023kondo,chou2023kondo,chen2023metallic,chen2024emergent,ye2024hopping,
ekahana2024anomalous,das2025moire} and less so on interactions via the lattice degrees of freedom \cite{ojajarvi2018competition,feng2020interplay,choi2021dichotomy}, which -- as will be shown in this work -- play a previously underappreciated role, at least at elevated temperatures.

Traditionally, the position of FBs has been inferred from spectroscopic probes, e.g., angle-resolved photoemission spectroscopy (ARPES) \cite{kang2020topological,liu2020orbital}, or extensive doping studies combined with low-temperature thermodynamic probes \cite{sales2021tuning,sales2022chemical}. However, both approaches can be experimentally demanding and time-consuming to perform across broad materials families and FBs above $E_\text{F}$ are typically inaccessible to ARPES. A simple method that relies only on analyzing electronic transport of a single material would be highly attractive, especially for rapidly screening new candidate systems.

Here, we show that a FB near $E_\mathrm{F}$ leaves a robust and universal fingerprint in the temperature dependence of the electrical resistivity at elevated temperatures owing to electron-phonon-mediated interband scattering. As phonons enable transitions from dispersive conduction bands into a nearby FB (Fig.\,\ref{Fig1}c), the drastic change in available scattering phase space generates a strongly energy-dependent carrier relaxation time (Fig.\,\ref{Fig1}d). This produces a distinct modification of the transport function $\sigma(E)$, which in turn yields characteristic deviations from the standard Bloch-Gr\"uneisen scaling at elevated temperatures. Most importantly, the sign of the curvature of $\rho(T)$ directly reveals whether $E_\mathrm{F}$ lies within or away from the FB and the temperature evolution of $d\rho/dT$ encodes information about the energy offset between $E_\text{F}$ and the FB edge, $\vert E_\text{F}-E_\text{edge} \vert$.

We developed a simple two-band model that captures these effects and show that its predictions do not depend qualitatively on microscopic details of the electron-phonon coupling by comparison with state-of-the-art \textit{ab initio} electron-phonon scattering calculations. Furthermore, we introduce a modified Bloch-Gr\"uneisen law, which produces quantitative agreement with transport measurements on a range of FB materials, such as \ce{Ni3(In,Sn)}, \ce{CoSn}, \ce{Ca(Rh,Ru)2}, \ce{Ni3Ge}, \ce{Pt3Sn}, and \ce{Pd3Pb}. Our work thus provides a broadly applicable framework for tracking FBs and enables rapid exploration and screening of promising candidate materials.

\section*{Results and Discussion}
\subsection*{Phenomenological model for interband scattering}
\begin{figure}[t]
\centering
 \includegraphics[width=0.48\textwidth]{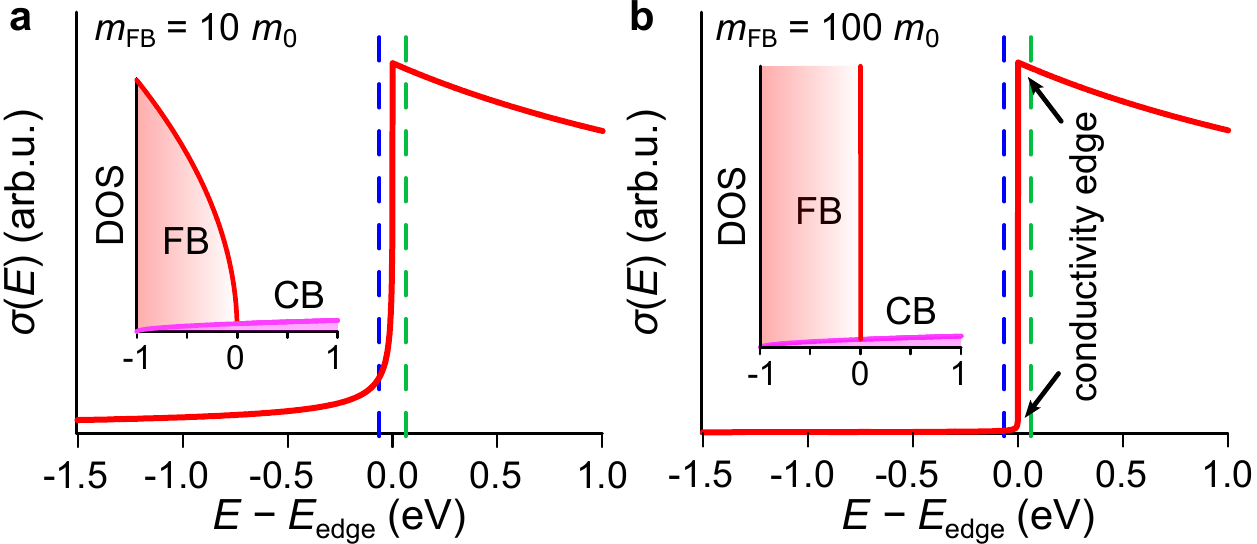}
\caption{\textbf{Energy-dependent transport function for interband scattering in our phenomenological two-band model.} \textbf{a}, Flat band (FB) with an effective mass $m_\text{FB}=10\,m_0$ and \textbf{b}, $m_\text{FB}=100\,m_0$. Effective mass ($m_\text{CB}=m_0$) and band filling ($n_\text{CB}= 1.6\times 10^{21}$\,cm$^{-3}$) of the dispersive conduction band (CB) are the same for \textbf{a} and \textbf{b}. Blue and green dashed lines correspond to $E_\text{F}-E_\text{edge}=-50$ and $+50$\,meV, respectively. Black arrows in \textbf{b} indicate a well defined edge in the spectral conductivity. The conductivity edge becomes smeared when the effective mass of the FB becomes smaller.
}
\label{Fig2}
\end{figure}

\noindent When FBs are present near $E_\text{F}$, the large electronic density of states (DOS) induces a rapid increase in the scattering phase space, even when charge carriers scatter off phonons, since the energy dependence of the scattering rate $\tau^{-1}_\text{ep}(E) \sim D(E)$ is directly proportional to the energy dependence of the DOS, $D(E)$, by Fermi's golden rule. This leads to a significant change of the carrier relaxation time $\tau_\text{ep}(E)$ and therefore the transport distribution function $\sigma(E)\sim \tau_\text{ep}(E)$ in the energy range where the FB overlaps with the more dispersive conduction bands (Fig.\,\ref{Fig1}d). To demonstrate this qualitatively, we have developed a two-band interband scattering model in the framework of semiclassical Boltzmann transport theory, which is inspired by Mott's early work on the resistivity and thermoelectric power of the transition metals \cite{mott1936resistance}. Our phenomenological model takes into account two parabolic bands with different effective masses, $m^*$: one parabolic band is highly dispersive ($m^* = m_0$) and the other parabolic band is much flatter ($m^* \gg m_0$), where $m_0$ is the free electron mass. The FB provides large additional phase space for phonon-mediated interband scattering processes, resulting in a distinct shape of $\tau_\text{ep}(E)$ and $\sigma(E)$ (Fig.\,\ref{Fig2}). We point out that the additional DOS in the overlapping region of the FB and CB does not contribute more conduction, but rather reduces the conductivity compared to the CB itself as scattering is enhanced. Details regarding the modeling framework are provided in the Supplementary Information (SI). Note that we do not include electron-electron scattering processes, for which phase space arguments would yield a $\tau^{-1}_\text{ee}(E) \sim D^2(E)$ scaling.

\begin{figure*}[tbh]
\centering
 \includegraphics[width=1\textwidth]{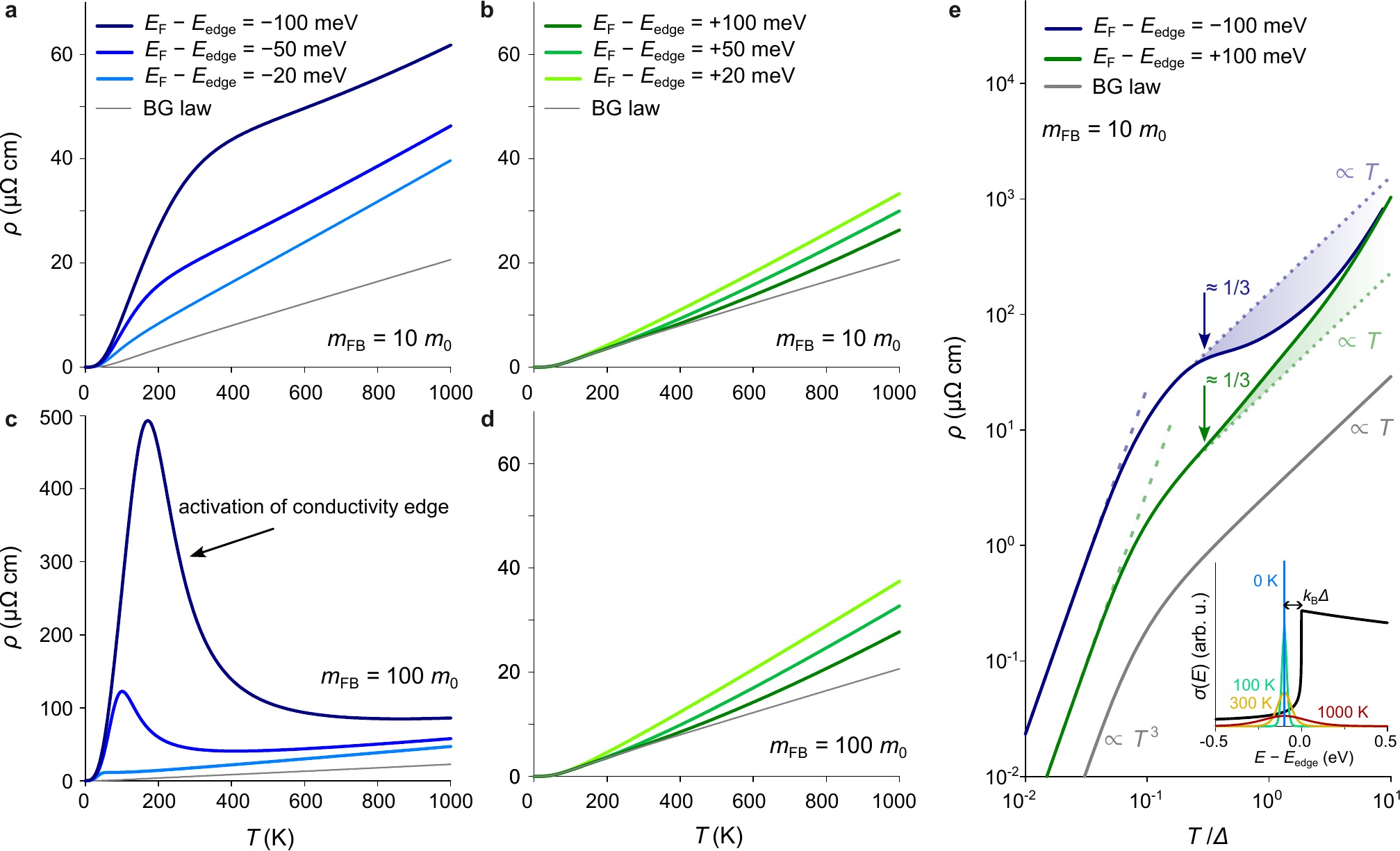}
\caption{\textbf{Temperature-dependent resistivity anomalies due to flat-band-driven interband scattering.} Corresponding to the transport functions shown in Fig.\,2, we calculated $\rho(T)$ for various positions of the FB edge: \textbf{a} and \textbf{c}, $E_\text{edge}$ sits above $E_\text{F}$ and \textbf{b} and \textbf{d}, below $E_\text{F}$. Calculations via the standard Bloch-Gr\"uneisen law (BG), which disregard any energy dependence of the phase space of carrier scattering are plotted for comparison (grey). \textbf{e}, Double logarithmic plot of $\rho(T)$ versus $T/\Delta$, where $k_\text{B}\Delta=\vert E_\text{F}-E_\text{edge}\vert$ is an energy scale associated with the Fermi level relative to the conductivity edge. Inset displays $\sigma(E)$ from Fig.\,2a together with the derivative of the Fermi-Dirac distribution function at different temperatures. Just above $T/\Delta \approx 0.33$, $\rho(T)$ decreases/increases compared to the BG scaling (grey) for $E_\text{F}$ below/above the FB edge. At even higher temperatures, $T/\Delta\gtrsim 10$, the behavior of $\rho(T)$ is heavily determined by $\sigma(E)$ contributions far from $E_\text{F}$, leading to almost identical scaling for $E_\text{F}<E_\text{edge}$ and $E_\text{F}>E_\text{edge}$.
}
\label{Fig3}
\end{figure*}


\begin{figure*}[tbh]
\centering
 \includegraphics[width=1\textwidth]{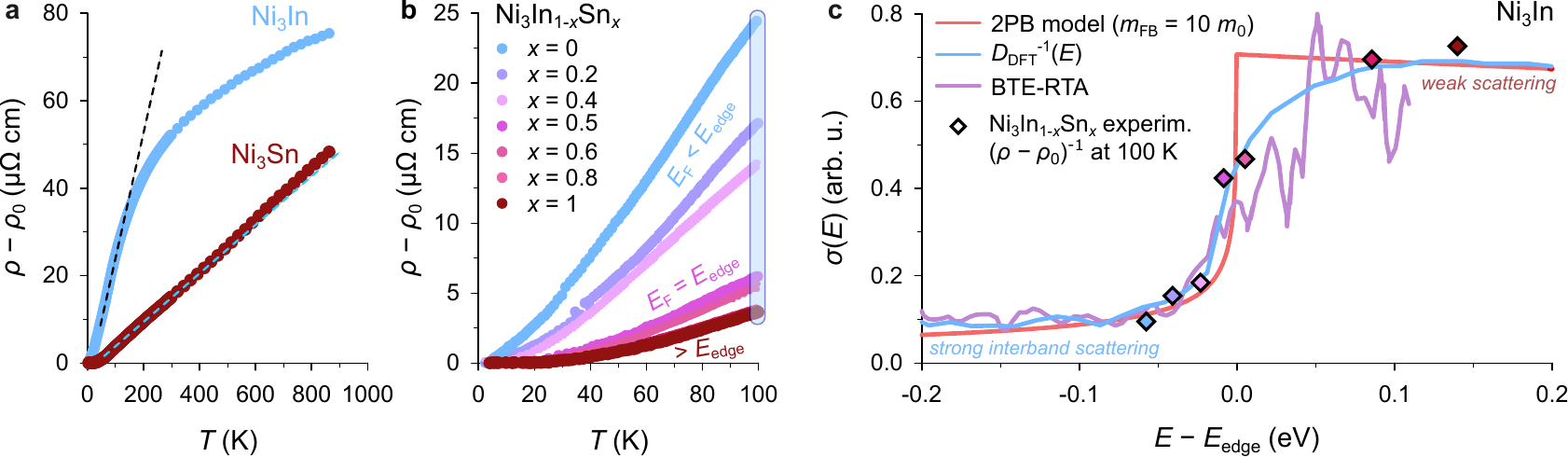}
\caption{\textbf{Tuning and reconstructing the energy-dependent transport function of the kagome metal \ce{Ni3In}.} \\\textbf{a}, Electrical resistivity of \ce{Ni3In} and \ce{Ni3Sn} in a broad temperature range 2\,--\,860\,K. At low temperatures, the slope of $\rho(T)$ varies by nearly an order of magnitude among the two compounds, but at high temperatures, when $k_\text{B}T$ becomes comparable to $|E_\text{F}-E_\text{edge}|$, the slopes converge, consistent with our phenomenological model (Fig.\,3). \textbf{b}, Temperature-dependent resistivity of \ce{Ni3In_{1-x}Sn_x}, which forms a solid solution in the entire composition range, allowing exceptional tunability of the FB states with respect to $E_\text{F}$. \textbf{c}, Semiclassical Boltzmann transport calculations using first-principles electron-phonon scattering rates within the relaxation time approximation (RTA) yield a step-like $\sigma(E)$ around the upper FB edge ($E_\text{edge}$). A similarly shaped $\sigma(E)$ is obtained from a simplified lifetime model based on the DFT density of states, $\tau(E)\propto D_\text{DFT}^{-1}(E)$. The two-parabolic-band model (see text and Methods) captures both the DOS enhancement and the associated drop in $\sigma(E)$ around $E_\text{edge}$. Experimentally, $\sigma(E)$ can be reconstructed by subtracting the residual resistivity to isolate intrinsic electron-phonon scattering contributions. The fact that the experimentally derived $\sigma(E)$ agrees well with the entirely DOS-based models (2PB, $D_\text{DFT}^{-1}$) highlights that microscopic details, such as group velocities and electron-phonon coupling, do not vary significantly.
}
\label{Fig4}
\end{figure*}

\begin{figure*}[tbh]
\centering
 \includegraphics[width=1\textwidth]{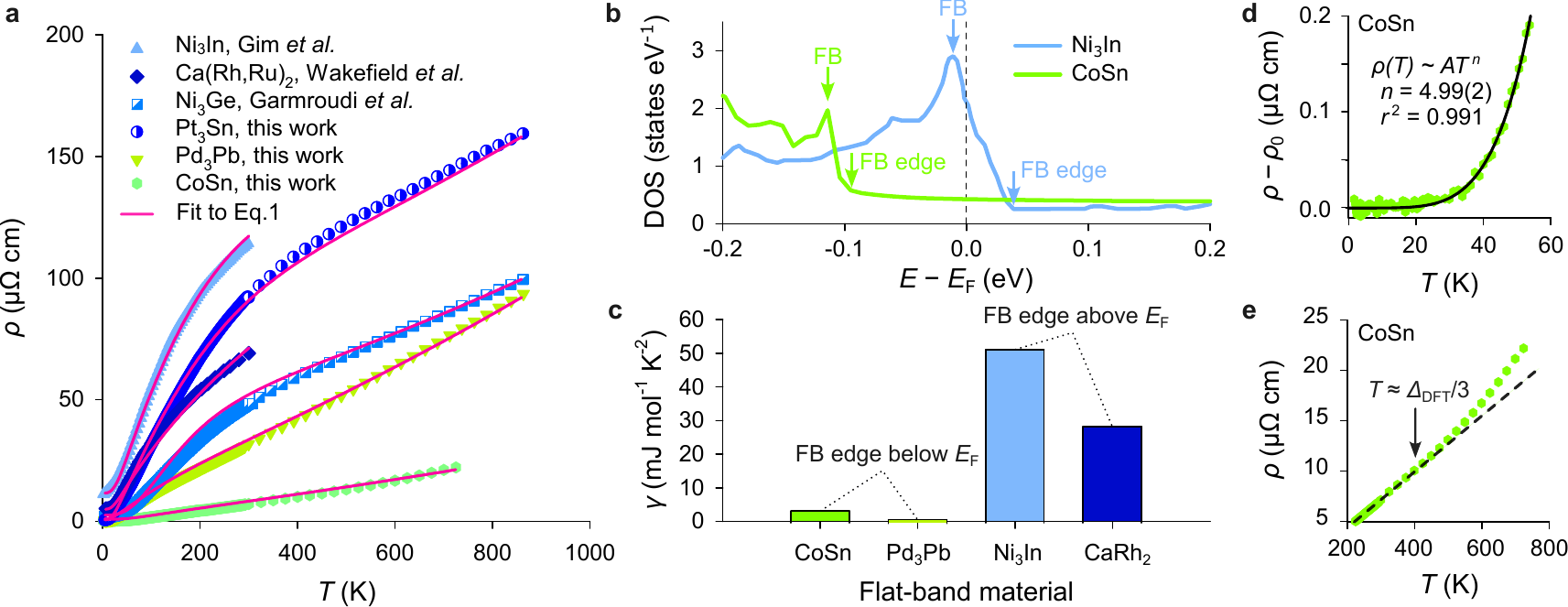}
\caption{\textbf{Universal description of resistivity in flat-band compounds by modified Bloch-Gr\"uneisen formula.} \\ \textbf{a}, Temperature-dependent electrical resistivity for several different FB compounds. Data for the kagome metal \ce{Ni3In}, the pyrochlore metal \ce{CaRh2} and the cubic compound \ce{Ni3Ge} were taken from refs.\,\cite{gim2023fingerprints}, \cite{wakefield2023three} and \cite{garmroudi2025energy}, respectively. Experimental data on polycrystalline samples of \ce{Pt3Sn} and \ce{Pd3Pb} (isostructural to \ce{Ni3Ge}) and on single crystals of the kagome metal CoSn, measured along the crystallographic $c$ direction were obtained in this work. Red solid lines show single-parameter fits, using a modified Bloch-Gr\"uneisen law (Eq.\,1) taking into account the energy-dependent scattering phase space via the DFT-derived DOS of all compounds. \textbf{b}, DFT-DOS of \ce{Ni3In} \cite{ye2024hopping} and CoSn (this work). For the former, the FB edge is expected to be slightly above $E_\text{F}$, while for the latter the FB is predicted to be entirely below $E_\text{F}$. \textbf{c}, Sommerfeld coefficient of specific heat \cite{ye2024hopping,wakefield2023three,meier2020flat}, being consistent with the notion established in \textbf{b}. \textbf{d}, Low-temperature resistivity in the temperature range 1.5\,--\,55\,K of CoSn (this work) can be well described by a power law $T^n$, with $n=4.99(2)$, in excellent agreement with electron-phonon intraband scattering. \textbf{e}, High-temperature resistivity deviates from linear behavior at around $T\approx 400\,\text{K}\approx\Delta_\text{DFT}/3$, hinting at the presence of a FB $\approx 0.1$\,eV below $E_\text{F}$.
}
\label{Fig5}
\end{figure*} 

Figure \ref{Fig3} shows the results of these calculations and their impact on the qualitative behavior of $\rho(T)$ for two different scenarios: (i) an effective mass of the FB $m_\text{FB}$ equaling $10\,m_0$ and (ii) $m_\text{FB}=100\,m_0$. For scenario (i), interband scattering results in a smeared edge in the energy-dependent transport distribution function, i.e. in the spectral conductivity $\sigma(E)$ (Fig.\,\ref{Fig2}a). Figure \ref{Fig3}a shows $\rho(T)$ when the Fermi level $E_\text{F}$ lies below the FB edge $E_\text{edge}$, where the conductivity $\sigma(E)$ is low. In contrast, Figure \ref{Fig3}b presents $\rho(T)$ for $E_\text{F}$ positioned above the FB edge, in the region of high $\sigma(E)$. It is obvious that even for the case where $m_\text{FB}=10\,m_0$, deviations from the standard Bloch-Gr\"uneisen (BG) law are observed (grey curve), which would yield a power law $\rho \propto T^n$ at low temperatures ($n=5$ for intraband and $n=3$ for interband scattering) transitioning into $\rho \propto T$ at high temperatures. Most importantly, there is a qualitative difference in the $\rho(T)$ behavior, namely a sublinear resistivity and a hump-like feature for $E_\text{F}-E_\text{edge} < 0$ (Fig.\,\ref{Fig3}a) and a superlinear $\rho(T)$ for $E_\text{F}-E_\text{edge} > 0$ (Fig.\,\ref{Fig3}b), both occurring at elevated temperatures where the BG law yields a linear slope. As $\vert E_\text{F}-E_\text{edge} \vert$ becomes smaller, these features shift towards lower temperatures. Note that the present calculations consider an almost fully filled FB with localized hole-type carriers. For an electron-like FB, which produces the exact same behavior, the situation would be reversed, i.e., sublinear (superlinear) $\rho(T)$ behavior would occur for $E_\text{F}$ above (below) the FB edge.

Increasing the effective mass of the FB up to $m_\text{FB}=100\,m_0$ results in an extremely asymmetric landscape of charge transport that is close to zero below $E_\text{edge}$ and jumps to a large value for $E>E_\text{edge}$ (Fig.\,\ref{Fig2}b). Examining $\rho(T)$, one finds that the hump has significantly sharped and now even forms a resistivity maximum with semiconductor-like behavior ($d\rho/dT < 0$) at high temperatures when $E_\text{F}-E_\text{edge} < 0$, despite the absence of a gap and the presence of a large DOS at $E_\text{F}$ (Fig.\,\ref{Fig3}c). In the present case, it is clear that this “anomalous" $\rho(T)$ stems from thermal activation of charge carriers across the scattering-induced edge in $\sigma(E)$ (Fig.\,\ref{Fig2}), eventually outweighing the increase of $\rho(T)$ due to an increase of the phonon amplitude with $k_\text{B}T$. In metals with a smooth and more energy-independent $\sigma(E)$, the latter dominates. Consequently, the sublinear $\rho(T)$ in Figure \ref{Fig3}a corresponds to the onset of this competition. As the negative derivative of the Fermi-Dirac distribution function ($-\partial f/\partial E$) broadens with temperature, sub- and superlinear $\rho(T)$ behaviors naturally emerge from either high or low $\sigma(E)$ regions becoming thermally activated. Figure \ref{Fig3}e shows a double logarithmic plot of $\rho(T)$ versus $T/\Delta$, where $k_\text{B}\Delta=\vert E_\text{F}-E_\text{edge}\vert$ represents the proximity of $E_\text{F}$ to the FB. At the lowest temperatures, $\rho(T)$ follows the BG scaling with $T^3$ or $T^5$. As $T$ reaches a fraction of $\Delta$ (about $\Delta/3$ or higher), strong deviations from the BG scaling are obvious (vertical arrows in Fig.\,\ref{Fig3}e). At $T \gg \Delta$, contributions of $\sigma(E)$ far from $E_\text{F}$ become important. Since $\sigma(E)$ decreases monotonically for $E\rightarrow \pm\infty$ in our model, the same scaling is obtained, irrespective of $E_\text{F}$. We have also considered a more realistic representation of the FB using a boxcar-type DOS to capture its finite bandwidth $W$, and the results are consistent with the discussion above in the case that $W>\Delta$ (see Fig.\,S1 and Methods for details).

\subsection*{Tracking flat bands in real materials}
\noindent Next, we evaluate how this simplified framework applies to real FB materials. In transition-metal compounds, FBs can originate from geometrically frustrated destructive interference patterns among subsets of the $d$ orbitals \cite{kang2020topological,liu2020orbital,ye2024hopping} or a lack of $d$-$d$ hopping across certain crystal directions \cite{ahn2018coexistence,garmroudi2025energy}. Additional dispersive bands typically intersect the Fermi level or the FB acquires dispersion along other Brillouin-zone directions. As a result, the DOS is better approximated by a step function (Fig.\,2) than by a narrow delta-like peak \cite{garmroudi2025topological}.

We start by analyzing the bilayer kagome metal \ce{Ni3In} -- a recently discovered FB compound that has drawn considerable interest for exhibiting strange-metal and correlated electron behavior at low temperatures \cite{ye2024hopping,gim2023fingerprints,han2024molecular,souza2025resolving,mahankali2025correlated}. Substitution of In with Sn is possible in a complete solid solution enabling exceptional tuning of $E_\text{F}$ relative to the FBs \cite{garmroudi2025topological}. Here, we investigate $\rho(T)$ of polycrystalline \ce{Ni3In_{1-x}Sn_x} across a wide composition range ($0\leq x\leq1$) and temperature interval ($2-860$\,K), allowing us to reconstruct the transport function $\sigma(E)$. First, we note that the stark contrast in $\rho(T)$ of \ce{Ni3In} and \ce{Ni3Sn} (Fig.\,\ref{Fig4}a) (even stronger in single crystals \cite{ye2024hopping}) is in excellent agreement with our interband scattering model (Fig.\,\ref{Fig2}), which predicts sub(super)linear behavior when $E_\text{F}<E_\text{edge}$ ($E_\text{F}>E_\text{edge}$) and a similar slope of $\rho(T)$ once $k_\text{B}T$ becomes comparable to $|E_\text{F}-E_\text{edge}|$. At low temperatures, the slope of $\rho(T)$ is several times larger for \ce{Ni3In} than in \ce{Ni3Sn}, and instead of a gradual evolution with Sn substitution, we find a rather abrupt change around $x=0.5$ (Fig.\,\ref{Fig4}b) that we ascribe to the transition of the FB edge through $E_\text{F}$. To obtain a quantitative measure of $\sigma(E)$, we extracted the residual resistivity $\rho_0$ from $\rho(T)$ and plotted the conductivity, $(\rho-\rho_0)^{-1}$, against $E - E_\text{edge}$ (Fig.\,\ref{Fig4}c). The mapping from chemical substitution $x$ to energy $E_\mathrm{F}-E_\mathrm{edge}$ is done via considering alloy-averaged DFT calculations \cite{garmroudi2025topological}. This reconstruction of $\sigma(E)$ reveals a clear step-like conductivity edge in \ce{Ni3In}, closely resembling the $\sigma(E)$ shape derived from our toy model (Fig.\,\ref{Fig2}a). To compare our experimental observations with theory, we computed $\sigma(E)$ within several interband-scattering frameworks of increasing complexity: (i) our two-parabolic band (2PB) model with $m_\text{FB}\approx 10\,m_\text{CB}$, (ii) a modified BG law that estimates the energy dependence of $\tau(E)$ via the DFT-DOS $\tau(E)\propto D_\text{DFT}^{-1}(E)$ (discussed in more detail later), and (iii) full \textit{ab initio} electron-phonon Boltzmann transport calculations within the relaxation-time approximation (RTA). The good agreement among these approaches both for $\sigma(E)$ at 100\,K (Fig.\,\ref{Fig4}c) and the calculated $\rho(T)$ at high $T$ (Fig.\,S2) indicates that electron-phonon scattering is not strongly influenced by the microscopic details of the electron and phonon dispersions or the specific coupling matrix elements, but is governed predominantly by the available phase space, i.e., the DOS. It is also interesting to note that the behavior of the experimental $\sigma(E)$ at 100\,K depends inversely on the DOS and does not follow a $D^{-2}(E)$ dependence, which would be expected in a Fermi liquid with dominant electron--electron scattering, where the scattering phase space is proportional to the DOS squared.  

Surprisingly, we find that electron-phonon interband scattering also provides an excellent description of the high-temperature resistivity in several other transition-metal-based FB candidate materials. Building on this, we propose a modified BG formula, which, once $D_\text{DFT}(E)$ and the Debye temprature $\Theta_\text{D}$ are known, requires only a single free fit parameter. Figure \ref{Fig5}a shows experimental electrical resistivity data for different FB compounds in a broad temperature range (2 up to 860\,K). Data for the kagome metal \ce{Ni3In}, the pyrochlore metal \ce{CaRh2} and \ce{Cu3Au}-type \ce{Ni3Ge} were taken from refs.\,\cite{gim2023fingerprints}, \cite{wakefield2023three} and \cite{garmroudi2025energy}, respectively. Polycrystalline samples of \ce{Cu3Au}-type \ce{Pt3Sn} and \ce{Pd3Pb}, predicted to host topological dispersive bands together with FBs around $E_\text{F}$ \cite{kim2017coexistence,ahn2018coexistence,khan2020fermi}, were synthesized via a two-step solid-liquid reaction and induction melting technique. We also synthesized and investigated single crystals of the kagome metal CoSn, which has previously been extensively studied at low temperatures \cite{kang2020topological,liu2020orbital,meier2020flat,sales2021tuning,huang2022flat} but not at high temperatures $T>300\,$K. Here, we collect $\rho(T)$ data of \ce{Pt3Sn}, \ce{Pd3Pb} and CoSn from 2\,--\,860\,K. From Figure \ref{Fig5}a, it is clear that sub- and superlinear resistivity trends, strikingly similar to those shown in Fig.\,\ref{Fig3}a and b and Fig.\,\ref{Fig4}a, are seen across these different intermetallic systems. To validate the above-described framework of electron-phonon interband scattering for all these different materials, we calculated $\rho(T)$ by extending the Bloch-Gr\"uneisen law with an energy-dependent scattering rate $\tau^{-1}_\text{ep}(E) \sim D_\text{DFT}(E)$ that takes the rapid variation of scattering-phase space induced by the FBs into account via the DFT-DOS for these compounds \cite{jain2013commentary,ye2024hopping,garmroudi2025energy,ahn2018coexistence}. This yields a modified BG formula given by

\begin{multline}
\label{fitformula}
    \rho(T)=C\left(\frac{T}{\Theta_\text{D}}\right)^n \int_0^{\Theta_\text{D}/T}\frac{t^n}{\left(e^t-1\right)\left(1-e^{-t}\right)}dt \\ \left[\int_{-\infty}^{\infty}D^{-1}_\text{DFT}(E)\left(-\frac{\partial f}{\partial E}\right)dE \right]^{-1}\,. 
\end{multline}

\noindent Here, $\Theta_\text{D}$ represents the Debye temperature, $n$ is an exponent which is either 5 for conventional \textit{intra}band scattering or 3 for \textit{inter}band scattering (important if FB states are present at $E_\text{F}$). $D_\text{DFT}(E)$ is the DFT-DOS and $C$ is a constant that includes various physical parameters, such as the electron-phonon coupling strength $\lambda_\text{ep}$ and carrier group velocities $v_\text{e}$. Since $C$ is difficult to estimate, we treated it as a temperature- and energy-independent fit parameter. We stress that this prefactor is the only free parameter in our modeling procedure. The red solid lines in Figure \ref{Fig5}a are single-parameter fits to Eq.\,\ref{fitformula}, yielding good agreement with all the FB materials we studied. 

Besides electron--phonon interband scattering, there are several other mechanisms that can lead to similar $\rho(T)$, especially sublinear/saturating behavior at elevated temperatures. In $f$-electron materials, for example, a flattening of $\rho(T)$ or even the appearance of a broad maximum is commonly associated with a crossover from coherent (Fermi-liquid) to incoherent Kondo scattering \cite{lavagna1982resistivity,sumiyama1986coherent}. In conventional semimetals or doped narrow-gap semiconductors, thermally activated carriers across a (pseudo)gap can likewise produce a flattening or maximum in $\rho(T)$ at elevated temperatures \cite{garmroudi2021boosting,garmroudi2022large}. Also, resistivity saturation at the Mott-Ioffe-Regel (MIR) limit can result in a reduced slope of $\rho(T)$ at high temperatures as the carrier mean free path reaches the interatomic distance \cite{gunnarsson2003colloquium}. In the present case, however, the investigated compounds are good metals (well below the MIR limit) whose transport properties are well-described within an electron-phonon scattering framework at high temperatures. This interpretation is further supported by state-of-the-art $\rho(T)$ calculations of \ce{Ni3In} (Fig.\,S2) and \ce{Ni3Ge} \cite{garmroudi2025energy} in the relaxation time approximation (RTA) leveraging the Phoebe code \cite{cepellotti2022phoebe}, where electron-phonon interactions and scattering rates were obtained directly from first principles. 
At this point, we also note that similar observations regarding the pivotal role of energy-dependent carrier scattering for the qualitative behavior of $\rho(T)$ have been made in other materials, such as binary transition metal alloys, Heusler compounds and skutterudites \cite{garmroudi2023high,fedorova2022anomalous,garmroudi2023pivotal,reumann2022thermoelectric,rogl2022understanding}. Lastly, using the previously discussed two-parabolic-band framework \cite{parzer2025seeband}, we derived a fit formula (Eq.\,7 in Methods) to pinpoint the position of the FB and determine its normalized effective mass without requiring any information on the DOS of the material. This is particularly useful when the DFT-calculated DOS is unavailable or unreliable, as is often the case for FBs.

While our phenomenological model assumes well-defined quasiparticles, strong electron--phonon scattering can lead to significant spectral broadening. We expect this broadening to smear the spectral conductivity edge, analogous to the behavior observed in transition-metal alloys exhibiting strong disorder-induced interband scattering \cite{pryga2025thermoelectric}. Furthermore, spectral broadening can facilitate Zener tunneling, a quantum transport mechanism that becomes particularly relevant when linewidths become comparable to the band offsets and the associated group velocities are large. Such effects can be captured within the Wigner transport formalism which takes into account the off-diagonal matrix element contributions to the velocity operator \cite{cepellotti2021interband}, as recently demonstrated for the kagome metal \ce{Ni3In} \cite{garmroudi2025topological}.

Figure \ref{Fig5}b compares the DFT-calculated DOS of two FB materials, that have received great interest for displaying FBs close to $E_\text{F}$: the kagome metals \ce{CoSn} \cite{kang2020topological} and \ce{Ni3In} \cite{ye2024hopping}. We note that DFT calculations employing standard GGA-PBE exchange correlation functionals \cite{perdew1996generalized} predict that the FB edge is above $E_\text{F}$ in \ce{Ni3In}, whereas it is more than 0.1\,eV below $E_\text{F}$ in \ce{CoSn}. This is in line with (i) the qualitative and quantitative resistivity behavior $\rho(T)$ at the lowest (Fig.\,\ref{Fig5}d) and also at the highest temperatures (Fig.\,\ref{Fig5}e, Fig.\,S1) and (ii) a small Sommerfeld coefficient of the specific heat $\gamma = 3.68\,$ mJ\,mol$^{-1}$\,K$^{-2}$ for CoSn \cite{meier2020flat}. Indeed, Figure \ref{Fig5}c shows that the Sommerfeld coefficient of CoSn \cite{meier2020flat} is more than an order of magnitude smaller than that observed for \ce{Ni3In} \cite{ye2024hopping}. This result implies that $E_\text{F}$ is located above the FB edge for CoSn. Similarly, we find that the FB is not placed right at but above $E_\text{F}$ in the topological semimetal \ce{Pd3Pb}, which has been theoretically predicted to display FB features at $E_\text{F}$ in DFT \cite{ahn2018coexistence}. Again, the very small $\gamma \approx 0.03\,$ mJ\,mol$^{-1}$\,K$^{-2}$ (Fig.\,\ref{Fig5}c and Fig.\,S4) is consistent with the lack of a sublinear $\rho(T)$. Instead, $\rho(T)$ displays superlinear behavior at high temperatures, consistent with phonon-mediated interband transitions changing the slope of $\rho(T)$ for $E_\text{F}>E_\text{edge}$ (Fig.\,\ref{Fig3}b).

Figure \ref{Fig5}d shows the low-temperature $\rho(T)$ subtracted by the residual resistivity $\rho_0$ for CoSn crystals measured along the crystallographic $c$ direction. A power law fit ($\rho-\rho_0 = AT^n$) in the temperature range 1.5\,--\,55\,K yields a resistivity exponent $n=4.99(2)$, which fits perfectly to what would be expected for conventional electron-phonon intraband scattering using the simple BG law. From the high-temperature resistivity in Fig.\,\ref{Fig5}e (and Fig.\,S3) it becomes clear, however, that phonon-mediated interband transitions into FB states below $E_\text{F}$ take place at elevated temperatures (cf. Fig.\,\ref{Fig3}e). Thus, external tuning parameters such as strain or doping are required to tune the FBs closer toward $E_\text{F}$. In contrast, for \ce{Ni3In}, where the FB edge lies above $E_\text{F}$, single crystals measured in the $ab$ (kagome) plane display strange-metal behavior with $n=1$ \cite{ye2024hopping}. Nevertheless, at elevated temperatures, $\rho_{ab}(T)$ can be well described by electron-phonon interband scattering taking into account our modified BG law (Eq.\,1). Lastly, we can rule out Kondo hybridization as the dominant mechanism governing the high-temperature resistivity behavior of \ce{Ni3In}, since the coherence temperature, $T^*\approx 50$\,K, inferred from recent Raman scattering measurements \cite{gim2023fingerprints}, is more than an order of magnitude lower than the temperature range over which $\rho(T)$ flattens.

\section*{Conclusion}
\noindent Using a phenomenological model, we reveal that phonon-mediated interband transitions, where charge carriers scatter from dispersive conduction bands into heavy (less mobile) FB states, give rise to characteristic temperature-dependent resistivity behaviors. If the Fermi level lies within the FB states, a sublinear $\rho(T)$ is found at elevated temperatures, whereas for $E_\text{F}$ away from the FB a superlinear $\rho(T)$ is observed. Within our description, these behaviors and their temperature regime depend sensitively on the position of the Fermi level relative to the FB edge, which enables tracking the FB states by fitting $\rho(T)$ of a given FB material. Our method provides an efficient way to determine the position of FB states compared to time-consuming doping studies or spectroscopic techniques such as ARPES. Most importantly, our work provides a surprisingly good description of the $\rho(T)$ behavior observed in a range of well-known intermetallic FB compounds studied in recent years and aligns with both DFT calculations and thermodynamic probes sensitive to FB states at $E_\text{F}$, such as low-temperature specific heat. By considering additional transport coefficients such as the thermoelectric power $S$, one can further infer whether the FB extends to higher or lower energies: for instance, $S < 0$ corresponds to the scenario depicted in Figure~\ref{Fig2}a and b, whereas $S > 0$ would indicate a FB extending towards $E>E_\text{F}$ \cite{garmroudi2023high,graziosi2024materials,garmroudi2025topological}. 

Since the proposed mechanism does not depend on the microscopic origin of the FB, but merely on the presence of a rapid change in the electronic density of states and the associated scattering phase space near $E_\text{F}$, it would be interesting to explore whether the present framework is applicable to moiré systems such as twisted bilayer graphene. Since moiré FBs also generate strong variations in the DOS, it is plausible to assume that similar effects may arise. However, the narrower bandwidth of moiré FBs may limit the magnitude of the resistivity anomaly within an electron--phonon scattering picture, as discussed in Methods, and, at the same time, also promote strong electron--electron interactions.

Summarizing, our work provides an alternative perspective on FB materials, one where electron-phonon as opposed to electron-electron scattering \cite{stewart1984heavy,schlottmann1989some,grenzebach2006transport} leads to salient features in the temperature dependence of the electrical resistivity at elevated temperatures, meriting further attention in other materials platforms with narrow bands as well.


%

\section*{Methods}
\label{methods}
\noindent Polycrystalline samples of the L1$_2$-ordered cubic compounds \ce{Pt3Sn} and \ce{Pd3Pb} (\ce{Cu3Au} structure) were synthesized via a two-step process. First, Pt and Sn as well as Pd and Pb were prereacted in an evacuated quartz tube at 1073\,K for 24\,h. Next, the resulting precursors were melted via high-frequency induction heating in a water-cooled copper cold boat under argon atmosphere. The samples were remelted several times for optimal homogeneity with a negligible relative mass loss of less than 0.05\,\%. Both samples were sealed in evacuated quartz ampoules and annealed at 1173\,K for one week, followed by furnace cooling. Single crystals of the kagome lattice compound CoSn were grown in Sn self flux, following the exact same procedure as described in ref.\,\cite{sales2021tuning}.

Rectangular bar-shaped samples of \ce{Pt3Sn} and \ce{Pd3Pb} were cut from the polycrystalline ingots using a high-speed cutting machine (Accutom by Struers) equipped with an aluminum oxide cutting wheel. Hexagonal bar-shaped samples obtained from the flux growth of CoSn were polished to obtain a rectangular cross section.

Low-temperature resistivity measurements were performed using an in-house setup at the Institute of Solid State Physics. Thin gold wires were spot-welded onto the sample surface in a four-probe geometry, the sample was mounted on a sample puck and the probe was immersed in a helium bath cryostat. Measurements are performed continuously as the sample space heats up slowly due to thermal contact with the environment. High-temperature measurements were performed in inert helium atmosphere using a commercially available setup (ZEM3 by ULVAC-RIKO) equipped with an infrared furnace that stabilizes the temperature. The setup employs the four-probe method to measure the high-temperature resistivity.

\subsection*{Two-parabolic band interband scattering model}

\noindent A simple, yet often quite accurate approximation for the energy dependence of the carrier relaxation time in the case of electron-phonon scattering is given by $\tau(E) \propto D^{-1}(E)$. This is particularly true for high temperatures where phonons with large momenta can scatter charge carriers into states all over the first Brillouin zone, whereas at low temperatures, momentum conservation may forbid certain scattering events involving large changes in the electron's momentum. In the case that two or more bands (acting as parallel transport channels) end up contributing to the transport properties, the single-band conductivity contributions have to be summed up $\sigma=\sum_i \sigma_i$.

We constructed a two-parabolic band interband scattering model to illustrate the effect of varying effective masses and Fermi level positions on the temperature-dependent resistivity behavior [Fig.\,3(a) to (d) in the main manuscript]. For parabolic bands, the integration of the carrier group velocities $v$ and relaxation times $\tau$ over $k$ space can be transformed into an integration over all energies by taking into account the spherical symmetry of the Fermi surface and introducing the density of states

\begin{equation}
D(E)=\frac{\left(2m^*\right)^{3/2}}{2\pi^2\hbar^3}\left(E-E_\text{edge}\right)^{1/2}\,.
\end{equation}

\noindent Here, $m^*$ denotes the effective mass and $E_\text{edge}$ the energy of the band edge for the parabolic band. Within Boltzmann transport theory, the temperature-dependent conductivity for a single parabolic band is defined as 

\begin{equation}
\label{cond}
\sigma(T)=\int_{-\infty}^{\infty}\tau(E)\,v^2(E)\,D(E)\,\left(-\frac{\partial f}{\partial E}\right)\,dE\,.
\end{equation}

\noindent The relaxation time is the inverse of the carrier scattering rate. For the latter, one can sum up the individual scattering rates due to individual scattering processes following Mathiessen`s rule $\tau^{-1}=\sum_i \tau^{-1}_i$. In a two-band model taking into account intra- and interband scattering processes and considering dominant electron-phonon scattering, this yields

\begin{equation}
\label{tau}
\tau(E)=\frac{\tau_{1\rightarrow1}(E)\tau_{1\rightarrow2}(E)}{\tau_{1\rightarrow1}(E)+\tau_{1\rightarrow2}(E)}\,,
\end{equation}

\noindent where the indices ${1\rightarrow1}$ and ${1\rightarrow2}$ represent intra- and interband scattering processes, respectively. The relaxation time due to intraband scattering $\tau_{1\rightarrow1}(E)\propto D_1^{-1}(E)$ is inversely proportional to the density of states of the first band and for interband scattering $\tau_{1\rightarrow2}(E)\propto D_2^{-1}(E)$ since the interband scattering phase space scales with $D_2(E)$.

In our model, we made the following approximation $\sigma_2=\sigma_\text{FB}\ll\sigma_1=\sigma_\text{CB}$. This is a very good approximation since we only consider very small values of $\vert E_\text{F}-E_\text{edge}\vert$, whereas the Fermi level and carrier concentration for the conduction band was fixed to a large (metallic) value of $n_\text{CB}\approx 1.6\times 10^{21}$\,cm$^{-3}$. Moreover, we consider an effective mass for the conduction band $m_\text{CB}=m_0$, where $m_0$ is the free electron mass, whereas holes in the FB are 10 and 100 times heavier for the two scenarios considered in Figure 2(a) and (d). We note that $\sigma$ scales inversely with the effective mass $m^*$ resulting in a 10 and 100 times weaker contribution of the FB from the higher effective mass alone.

Since $v^2(E)$ and $D(E)$ of the conduction band become weakly energy-dependent in the metallic regime ($E_\text{F}$ being located far from the band edge inside the conduction band), one only needs to integrate the energy dependence of $\tau(E)$ in \autoref{cond}, which is given by \autoref{tau} and simplifies to $\tau(E)\propto D_\text{tot}^{-1}(E)$ when the transition matrix elements for $\tau_{1\rightarrow1}$ and $\tau_{1\rightarrow2}$ are comparable. Here, $D_\text{tot}$ represents the total density of states, i.e. the sum of the FB and the dispersive-band DOS. 

The proportionality factor including e.g. the scattering matrix element for transitions between intial and final states $M_{i\rightarrow f}$ introduces an additional temperature dependence besides the one expected from solving \autoref{cond}.
At high temperatures, an additional linear temperature dependence in the proportionality factor of $\tau^{-1}(E,T)$ occurs due to an increase of the phonon oscillation amplitude. At low temperatures, where only phonons around the $\Gamma$ point with small wavevectors are active, the well-known Bloch-Gr\"uneisen (BG) law modifies the temperature dependence. In our model, we used the BG law as the energy-independent proportionality factor $\widetilde{\tau}^{-1}(T)$ in $\tau^{-1}(E,T)=\widetilde{\tau}^{-1}(T)\,D_\text{tot}(E)$.

This allows one to derive a modified BG formula, which takes into account the energy dependence of the carrier scattering phase space, which can become dramatically modified if FBs are present near the Fermi level. We therefore can write 

\begin{multline}
\label{modBG1}
    \rho(T)=\\\widetilde{\tau}^{-1}_\text{BG}\left[\int_{-\infty}^{\infty}\frac{1}{D_1(E)+D_2(E)}\left(-\frac{\partial f}{\partial E}\right)dE\right]^{-1},
\end{multline}

\noindent with $D_1(E)$ and $D_2(E)$ being the density of states of the dispersive and the FB, respectively. For simplification and assuming that the dispersive band is close to half filling and the FB is not, one can very reasonably assume that the energy dependence of $D_1(E)$ can be neglected $D_1(E)\approx D_1$ compared to $D_2(E)$. Thus, \autoref{modBG1} can be written as

\begin{multline}
\label{modBG2}
    \rho(T)=C\left(\frac{T}{\Theta_\text{D}}\right)^n \int_0^{\Theta_\text{D}/T}\frac{t^n}{\left(e^t-1\right)\left(1-e^{-t}\right)}dt \\ \left[\int_{-\infty}^{\infty}\frac{1}{D_1+m_2^{3/2}\left(E-E_\text{edge}\right)^{1/2}}\left(-\frac{\partial f}{\partial E}\right)dE\right]^{-1},
\end{multline}

\noindent assuming a parabolic band dispersion for the FB edge. Note that natural constants in the prefactors of $D_1$ and $D_2$ have been included in a constant global prefactor $C$. Defining the energy of the FB edge as the reference point $E_\text{edge}=0$ as well as rewriting the integrand of the second integral, one gets

\begin{multline}
\label{modBG3}
    \rho(T)=C\left(\frac{T}{\Theta_\text{D}}\right)^n \int_0^{\Theta_\text{D}/T}\frac{t^n}{\left(e^t-1\right)\left(1-e^{-t}\right)}dt\\ \left[\int_{-\infty}^{\infty}\frac{1}{1+\frac{m_2^{3/2}}{D_1}\left(E-E_\text{edge}\right)^{1/2}}\left(-\frac{\partial f}{\partial E}\right)dE\right]^{-1}.
\end{multline}

\noindent By fitting experimental resistivity data, measured in a broad temperature range, using \autoref{modBG3}, one can deduce information about the distance of the FB edge with respect to the Fermi level. If the Debye temperature $\Theta_\text{D}$ is known, e.g. from specific heat measurements, this leaves only three independent fit parameters: (A) The constant global prefactor $C$, (B) the relative contribution of the FB to the total density of states around the Fermi level given by the ratio $m_2^{3/2}/{D_1}$ and (C) the position of the Fermi level $E_\text{F}$ relative to the band edge, occuring in the derivative of the Fermi-Dirac distribution function $-\frac{\partial f}{\partial E}$. In our calculations of the electrical resistivity in Fig.\,3 of the main manuscript, the Debye temperature was set to a typical value of 400\,K and a low-temperature exponent $n=3$ for interband scattering was assumed.

\subsection*{Two-band model with finite bandwidth for the flat band}
\noindent Another interesting question is how $\rho(T)$ evolves when the FB is no longer modeled within a parabolic-band framework, which captures the steep DOS increase but fails to represent its confined bandwidth $W_\text{FB}$. To address this, we also calculated $\rho(T)$ for a corresponding two-band model where the FB is represented by a boxcar-type DOS: $D(E)=D_\text{FB}=\text{const.}$ within an interval of width $W_\text{FB}$ and $D(E)=0$ elsewhere (Fig.\,S1a). We conserve the total FB spectral weight by increasing $D_\text{FB}$ in proportion to the reduction of $W_\text{FB}$, which mimicks a realistic tuning of bandwidth by pressure or strain, where the number of states and orbitals remain fixed.

For small $W_\text{FB}$, this model yields a local suppression of $\sigma(E)$ (Fig.\,S1b), whereas larger $W_\text{FB}$ produces a step-like $\sigma(E)$ similar to the 2PB case, sketched in Fig.\,2. Figure S1c shows that for $E_\text{F}$ doped 0.1\,eV into the FB, reducing $W_\text{FB}$ initially increases $\rho(T)$ and produces a local maximum, similar to the trend obtained in the 2PB model with $m_\text{FB}= 100\,m_0$ (Fig.\,3c). Once $W_\text{FB}$ becomes small compared to $E_\text{F}$, however, this maximum disappears and re-emerges only at low temperatures when $E_\text{F}$ is shifted correspondingly (Fig.\,S1d). Additionally, the absolute value of the maximum remains lower for smaller $W_\text{FB}$ than for larger $W_\text{FB}$ (cf. Fig.\,S1c,d).

These results highlight that two parameters control both the presence and magnitude of the local maximum $\rho_\text{max}$ as well as the overall shape of $\rho(T)$:
(i) the distance of $E_\text{F}$ from the band edge (larger distances yield larger $\rho_\text{max}$), and (ii) the relative step size in $\sigma(E)$, i.e. difference between low- and high conductivity regions. Most importantly, a FB whose bandwidth is narrow relative to the thermal energy scale $k_\text{B}T$ cannot produce a sizeable anomaly within a purely electron-phonon scattering scheme, regardless of the position of $E_\text{F}$, because it just "burns" a narrow hole in $\sigma(E)$.

\subsection*{Ab initio electron-phonon transport calculations}
\noindent Electronic transport properties were computed within the semiclassical Boltzmann transport formalism. The calculations were performed using the Phoebe code, an open-source implementation for solving the electron and phonon Boltzmann transport equations \cite{cepellotti2022phoebe}.

Electron-phonon coupling matrix elements were obtained from density functional theory calculations using the JDFTx package, employing GBRV pseudopotentials and the Perdew–Burke-Ernzerhof (PBE) exchange-correlation functional. Wannier interpolation was used to efficiently evaluate the electron-phonon matrix elements on dense Brillouin-zone grids.

The electron-phonon scattering rates were calculated in the relaxation time approximation as

\begin{multline}
\frac{1}{\tau_{k m}} =
\frac{2\pi}{V N_k}
\sum_{m'\textbf{k}',\,\nu \textbf{q}}
\left| g_{mm'\nu}(\textbf{k},\textbf{k}') \right|^2 \\
\times [
(1 - f_{\textbf{k}'m'} + n_{\textbf{q}\nu})
\, \delta(\varepsilon_{\textbf{k} m} - \varepsilon_{\textbf{k}' m'} - \hbar \omega_{\textbf{q}\nu}) \\
+(f_{\textbf{k}'m'} + n_{\textbf{q}\nu})
\, \delta(\varepsilon_{\textbf{k} m} - \varepsilon_{\textbf{k}' m'} + \hbar \omega_{\textbf{q}\nu})]\delta\left(\textbf{k} - \textbf{k}' + \textbf{q}\right),
\end{multline}

where $m$ and $m'$ denote electronic band indices, $\nu$ labels the phonon mode,
$k$ and $k'$ are electron wave vectors, and $q = k' - k$ is the phonon wave vector.
The quantities $\varepsilon_{k m}$ and $\hbar \omega_{q \nu}$ are the electron and
phonon energies, respectively, while $f_{k m}$ and $n_{q \nu}$ are the
Fermi--Dirac and Bose--Einstein occupation factors. $N_k$ is the number of
$k$-points used in the Brillouin-zone sampling.

From the momentum- and band-resolved relaxation times $\tau_{k m}$, the
electrical conductivity tensor was computed as

\begin{equation}
\left[\rho\right]^{-1}_{\alpha\beta} =
\frac{e^2}{V N_k}
\sum_{\textbf{k} m}
\left(
- \frac{\partial f_{\textbf{k} m}}{\partial \varepsilon_{\textbf{k} m}}
\right)
\left(v_{\alpha} v_{\beta}\right)_{\textbf{k} m}
\tau_{\textbf{k} m},
\end{equation}
where $V$ is the unit-cell volume and $v_\alpha$ and $v_\beta$ are the alpha and beta components of the group velocity, respectively.

Transport integrals were evaluated using dense $k$-point meshes (typically
$75 \times 75 \times 75$) together with adaptive smearing to approximate the
Dirac delta functions. Only electronic states within a narrow energy window
around the Fermi level were included, controlled by a threshold on the
derivative of the Fermi--Dirac distribution, corresponding to a Fermi window
width of $\partial f / \partial T = 1 \times 10^{-10}$.

\subsection*{Data Availability} 
\noindent All data are available within the
article and its Supplementary Information. 
\vspace{0.3cm}
\subsection*{Code Availability} 
\noindent The codes underlying the figures presented in this article are available
from the corresponding authors upon request. 
\subsection*{Acknowledgements} 
\noindent The authors acknowledge Antoine Georges as well as the collaborators of the SuperC project for fruitful and stimulating discussions. Work at Los Alamos National Laboratory was performed under the auspices of the U.\,S. Department of Energy, Office of Basic Energy Sciences, Division of Materials Science and Engineering.
\subsection*{Funding} 
\noindent F.\,G. acknowledges a
Director’s Postdoctoral Fellowship through the
Laboratory and Directed Research \& Development (LDRD) program. S.\,M.\,T. acknowledges support from the Los Alamos LDRD program. X.\,Y. and S.\,P. acknowledge funding from the European Research Council (ERC AdG 101055088-CorMeTop). A.\,P. acknowledges support by the European Research Council (ERC Consolidator Grant 101231553-
METHEL). The Flatiron Institute is a division of the Simons Foundation.
\subsection*{Author contributions} 
\noindent F.\,G. synthesized \ce{Pt3Sn} and \ce{Pd3Pb} and X.\,Y. grew CoSn single crystals. F.\,G. measured the electrical resistivity of \ce{Pt3Sn}, \ce{Pd3Pb}, \ce{Ni3In_{1-x}Sn_x} ($x=0$, 0.2, 0.5, 0.8, 1) and, together with C.\,F., the resistivity of CoSn. I.\,S. measured the resistivity of \ce{Ni3In_{1-x}Sn_x} ($x=0.4$, 0.6). F.\,G. and P.\,F.\,S.\,R. performed specific heat measurements for \ce{Pd3Pb}. F.\,G. and A.\,P. developed the phenomenological model. J.\,C. performed ab initio electron--phonon transport calculations of \ce{Ni3In} using the Phoebe code. F.\,G. conceptualized the work and wrote the initial draft. S.\,P., S.\,M.\,T., E.\,D.\,B., A.\,P. and P.\,F.\,S.\,R. supervised the work. All authors discussed the results and edited the manuscript.

\subsection*{Competing interests} 
\noindent Priscila F.\,S. Rosa is an Editorial Board Member for Communications Materials but was not involved in the editorial review or peer review, nor in the decision to publish this article.

\subsection*{Figure captions}

\noindent \textbf{Fig.1 $\vert$ Sketch of flat band formation and phonon-mediated interband scattering.} Localized states and flat bands can arise in \textbf{a}, materials with localized atomic orbitals and \textbf{b}, driven by destructive phase interference patterns. \textbf{c}, Sketch of phonon-mediated interband scattering between a dispersive and a flattened band. \textbf{d}, Schematic density of states for a metallic system with a dispersive conduction band and overlapping FB. The variation of scattering phase space creates a sharp drop in the carrier relaxation time (schematically drawn as blue line).

\noindent \textbf{Fig.2 $\vert$ Energy-dependent transport function for interband scattering in our phenomenological two-band model.} \textbf{a}, Flat band (FB) with an effective mass $m_\text{FB}=10\,m_0$ and \textbf{b}, $m_\text{FB}=100\,m_0$. Effective mass ($m_\text{CB}=m_0$) and band filling ($n_\text{CB}= 1.6\times 10^{21}$\,cm$^{-3}$) of the dispersive conduction band (CB) are the same for \textbf{a} and \textbf{b}. Blue and green dashed lines correspond to $E_\text{F}-E_\text{edge}=-50$ and $+50$\,meV, respectively. Black arrows in \textbf{b} indicate a well defined edge in the spectral conductivity. The conductivity edge becomes smeared when the effective mass of the FB becomes smaller.

\noindent \textbf{Fig.3 $\vert$ Temperature-dependent resistivity anomalies due to flat-band-driven interband scattering.} Corresponding to the transport functions shown in Fig.\,2, we calculated $\rho(T)$ for various positions of the FB edge: \textbf{a} and \textbf{c}, $E_\text{edge}$ sits above $E_\text{F}$ and \textbf{b} and \textbf{d}, below $E_\text{F}$. Calculations via the standard Bloch-Gr\"uneisen law (BG), which disregard any energy dependence of the phase space of carrier scattering are plotted for comparison (grey). \textbf{e}, Double logarithmic plot of $\rho(T)$ versus $T/\Delta$, where $k_\text{B}\Delta=\vert E_\text{F}-E_\text{edge}\vert$ is an energy scale associated with the Fermi level relative to the conductivity edge. Inset displays $\sigma(E)$ from Fig.\,2a together with the derivative of the Fermi-Dirac distribution function at different temperatures. Just above $T/\Delta \approx 0.33$, $\rho(T)$ decreases/increases compared to the BG scaling (grey) for $E_\text{F}$ below/above the FB edge. At even higher temperatures, $T/\Delta\gtrsim 10$, the behavior of $\rho(T)$ is heavily determined by $\sigma(E)$ contributions far from $E_\text{F}$, leading to almost identical scaling for $E_\text{F}<E_\text{edge}$ and $E_\text{F}>E_\text{edge}$.

\noindent \textbf{Fig.4 $\vert$ Tuning and reconstructing the energy-dependent transport function of the kagome metal \ce{Ni3In}.} \\\textbf{a}, Electrical resistivity of \ce{Ni3In} and \ce{Ni3Sn} in a broad temperature range 2\,--\,860\,K. At low temperatures, the slope of $\rho(T)$ varies by nearly an order of magnitude among the two compounds, but at high temperatures, when $k_\text{B}T$ becomes comparable to $|E_\text{F}-E_\text{edge}|$, the slopes converge, consistent with our phenomenological model (Fig.\,3). \textbf{b}, Temperature-dependent resistivity of \ce{Ni3In_{1-x}Sn_x}, which forms a solid solution in the entire composition range, allowing exceptional tunability of the FB states with respect to $E_\text{F}$. \textbf{c}, Semiclassical Boltzmann transport calculations using first-principles electron-phonon scattering rates within the relaxation time approximation (RTA) yield a step-like $\sigma(E)$ around the upper FB edge ($E_\text{edge}$). A similarly shaped $\sigma(E)$ is obtained from a simplified lifetime model based on the DFT density of states, $\tau(E)\propto D_\text{DFT}^{-1}(E)$. The two-parabolic-band model (see text and Methods) captures both the DOS enhancement and the associated drop in $\sigma(E)$ around $E_\text{edge}$. Experimentally, $\sigma(E)$ can be reconstructed by subtracting the residual resistivity to isolate intrinsic electron-phonon scattering contributions. The fact that the experimentally derived $\sigma(E)$ agrees well with the entirely DOS-based models (2PB, $D_\text{DFT}^{-1}$) highlights that microscopic details, such as group velocities and electron-phonon coupling, do not vary significantly.

\noindent \textbf{Fig.5 $\vert$ Universal description of resistivity in flat-band compounds by modified Bloch-Gr\"uneisen formula.} \\ \textbf{a}, Temperature-dependent electrical resistivity for several different FB compounds. Data for the kagome metal \ce{Ni3In}, the pyrochlore metal \ce{CaRh2} and the cubic compound \ce{Ni3Ge} were taken from refs.\,\cite{gim2023fingerprints}, \cite{wakefield2023three} and \cite{garmroudi2025energy}, respectively. Experimental data on polycrystalline samples of \ce{Pt3Sn} and \ce{Pd3Pb} (isostructural to \ce{Ni3Ge}) and on single crystals of the kagome metal CoSn, measured along the crystallographic $c$ direction were obtained in this work. Red solid lines show single-parameter fits, using a modified Bloch-Gr\"uneisen law (Eq.\,1) taking into account the energy-dependent scattering phase space via the DFT-derived DOS of all compounds. \textbf{b}, DFT-DOS of \ce{Ni3In} \cite{ye2024hopping} and CoSn (this work). For the former, the FB edge is expected to be slightly above $E_\text{F}$, while for the latter the FB is predicted to be entirely below $E_\text{F}$. \textbf{c}, Sommerfeld coefficient of specific heat \cite{ye2024hopping,wakefield2023three,meier2020flat}, being consistent with the notion established in \textbf{b}. \textbf{d}, Low-temperature resistivity in the temperature range 1.5\,--\,55\,K of CoSn (this work) can be well described by a power law $T^n$, with $n=4.99(2)$, in excellent agreement with electron-phonon intraband scattering. \textbf{e}, High-temperature resistivity deviates from linear behavior at around $T\approx 400\,\text{K}\approx\Delta_\text{DFT}/3$, hinting at the presence of a FB $\approx 0.1$\,eV below $E_\text{F}$.

\end{document}